\documentclass[11pt,preprint]{aastex}  
							
\usepackage{geometry}
\usepackage{longtable}

\usepackage{amsmath,amssymb,graphicx,fancyhdr,natbib, wasysym}

\usepackage[center,tiny]{titlesec}

\begin{document}

\section*{Compaction and Melt Transport in Ammonia-Rich Ice Shells: Implications for the Evolution of Triton}
\vspace{0.25in}
\centerline{\textbf{N.~P. Hammond$^1$, E.~M. Parmentier$^2$, A.~C. Barr$^3$}}

\begin{itemize}
\item[$^1$]N. P. Hammond, Centre for Planetary Sciences, University of Toronto at Scarborough, 1265 Military Trail, Toronto ON Canada. (noah.hammond@utoronto.ca)
\item[$^2$]E. M. Parmentier, Department of Earth, Environmental and Planetary Sciences, Brown University, 324 Brook Street, Box 1846, Providence RI 02912, USA. (e.m.parmentier@brown.edu)
\item[$^3$]A. C. Barr, Planetary Science Institute, 1700 East Fort Lowell, Suite 106, Tuscon, AZ 85719-2395, USA. (amy@psi.edu)
\item[]Accepted for publication in Journal of Geophysical Research - Planets. Copyright 2018 American Geophysical Union. Further reproduction or electronic distribution is not permitted.
\item[]In press, 2018 (2018JE005781)
\end{itemize}

\clearpage
\vspace{0.5in}
\baselineskip=15 pt
\noindent \textbf{Abstract}:  Ammonia, if present in the ice shells of icy satellites, could lower the temperature for the onset of melting to 176 K and create a large temperature range where partial melt is thermally stable. 
The evolution of regions of ammonia-rich partial melt could strongly influence the geological and thermal evolution of icy bodies.
For melt to be extracted from partially molten regions, the surrounding solid matrix must deform and compact. 
Whether ammonia-rich melts sink to the subsurface ocean or become frozen into the ice shell depends on the compaction rate and thermal evolution.
Here we construct a model for the compaction and thermal evolution of a partially molten, ammonia-rich ice shell in a one-dimensional geometry. 
We model the thickening of an initially thin ice shell above an ocean with $10\%$ ammonia. 
We find that ammonia-rich melts can freeze into the upper $5$ to $10$ kilometers of the ice shell, when ice shell thickening is rapid compared to the compaction rate. 
The trapping of near-surface volatiles suggests that, upon reheating of the ice shell, eutectic melting events are possible.
However, as the ice shell thickening rate decreases, ammonia-rich melt is efficiently excluded from the ice shell and the bulk of the ice shell is pure water ice.
We apply our results to the thermal evolution of Neptune's moon Triton. 
As Triton's ice shell thickens, the gradual increase of ammonia concentration in Triton's subsurface ocean helps to prevent freezing and increases the predicted final ocean thickness by up to $50$ km.

\baselineskip=20pt   
\parskip=2pt	

\section{Introduction}
Our solar system is overflowing with ocean worlds. 
Many icy satellites and large Kuiper Belt objects are thought to harbor subsurface oceans of liquid water beneath ice shells that are $10$s - $100$s of kilometers thick \citep{khurana1998induced, kivelson2002permanent, hussmann2006subsurface, postberg2009sodium, Hammond2013, nimmo2016reorientation}. Potential regions of partial melt within the ice shells could be important in influencing the geological and thermal evolution of ocean worlds. 

On several icy satellites there is strong evidence of melt being present near the surface, such as plumes of water vapor erupting from the tiger stripes of Enceladus \citep{postberg2009sodium,porco2014geysers}, and chaos terrains on Europa, which likely form in response to melting occurring only $1-2$ km below the surface \citep{head1999brine,sotin2002europa}. 
Additionally, some satellites show potential evidence for ``cryovolcanic'' terrains. These terrains are suggested to form when water-rich fluids and slurries erupt onto and flow across the surface \citep{kargel1995cryovolcanism, schenk1991fluid, fagents2003considerations, desch2009thermal}. 
Near-surface liquids on icy satellites could be sourced directly from the subsurface ocean, as is likely the case for Enceladus, or could be generated from within the ice shell by tidal heating. 

Understanding the dynamics of how melt is generated and migrates through the ice shell is crucial to explaining observations of near-surface water. Water should be negatively buoyant in the ice shell and should sink unless there are significant pressure gradients driving flow, such as from topography or ocean pressurization (e.g., \citealt{fagents2003considerations, showman2004resurfacing, manga2007pressurized}). 

The presence of chemical impurities in the ice shell could strongly influence the dynamics of melting on icy satellites. A wide variety of chemical compounds may be present in icy satellite ice shells, ranging in abundance from $0.1-15\%$ mass fraction \citep{mckinnon2008structure}, including CH$_4$, CO, CO$_2$, MgSO$_4$, NaCl, and NH$_3$ \citep{kargel1992ammonia}. Salts and other volatiles could allow for the generation of low-temperature eutectic melts, which has been suggested as a mechanism for generating near-surface melting on Europa \citep{pappalardo2004origin, han2005thermo, schmidt2011active}. 

Ammonia, in particular has the potential to strongly influence the evolution of icy satellites and Kuiper belt objects \citep{desch2009thermal}. 
The presence of ammonia can lower the solidus of the ammonia-water system to 176 K \citep{LK2002}, and ammonia is one of the only dissolved volatiles that reduces the density of the melt \citep{croft1988equation}. At the eutectic temperature, the density of the ammonia-rich melt is $\rho_l=0.946$ g$/$cm$^3$. The lower density increases the potential for positive buoyancy and cryovolcanism. 

Ammonia is thought to be most abundant in the outer solar system, where temperatures during planetary formation favored the condensation of ammonia from the solar nebula \citep{lewis1972low,lodders2003solar,dodson2009ice}. Ammonia has been detected spectroscopically on the surface of Pluto's moon Charon \citep{brown2000evidence}, on Uranus' satellite Miranda \citep{bauer2002near} and has also been measured in the plumes erupting on Enceladus \citep{waite2009liquid}. Frozen ammonia-solutions could also be present on the surfaces of other icy bodies in the solar system but may be difficult to detect. Due to damage from solar radiation, ammonia's spectroscopic signature could be erased after $10^4$ years in the Saturnian system and $10^9$ years in the Pluto system \citep{moore2007ammonia}. 

The amount of ammonia incorporated into icy bodies is uncertain. 
Early nebula condensation models by \citet{lewis1972low} suggested an NH$_3$ to H$_2$O mass ratio of $15 - 30\%$ in the outer solar system.
More recent models suggest a smaller amount of ammonia ranging from $\sim1$ to $15\%$ \citep{hersant2004enrichment, dodson2009ice}, with much of the uncertainty resulting from whether nitrogen was present in the gaseous nebula as N$_2$ or NH$_3$ \citep{hersant2004enrichment}.
Here we assume modest amounts of ammonia, in the range of $5 - 10$ wt. $\%$ in the ice shell. 
It should also be noted that under certain conditions aqueous ammonia can be consumed in reactions with other species, such as CO$_2$, and precipitate as a solid \citep{kargel1992ammonia, marion2012modeling}. 
In this work we assume ammonia is in aqueous form and do not treat chemical reactions with other species.

Although there are numerous observations of volatiles on the surface of icy bodies \citep{mccord1998salts, mccord1998non, brown2000evidence, dalton2005spectral}, the amount of volatiles in the bulk of their ice shells is uncertain. It is thought that the majority of volatiles partition in the subsurface ocean during ice shell formation and that the ice shell should become depleted in volatiles over time \citep{pappalardo2004origin}. The bulk of icy satellite ice shells, therefore, could be pristine water ice. If volatiles are not present within the ice shell, eutectic melting at low temperatures may not be possible.

A model that examines how ammonia-rich melts migrate and evolve during ice shell formation is necessary to predict the likely distribution of ammonia within an ice shell. Several investigators have explored the role of ammonia in cryovolcanic resurfacing during the thermal evolution of icy bodies \citep{stevenson1982volcanism, kargel1991rheological, kargel1995cryovolcanism, hogenboom1997ammonia, desch2009thermal, desch2017differentiation}. Perhaps the most detailed investigation of compaction and melt migration in icy satellite ice shells is by \citet{kalousova2014ice}. Using a one-dimensional melt migration model they investigated the downward propagation of melt on Europa from the near-surface to the subsurface ocean and looked at the timescale over which melt would be present in the ice shell. We take a similar approach, constructing a model based on studies of compaction in terrestrial magma chambers \citep{shirley1986compaction}. Studies of melt transport in terrestrial magma chambers provide valuable insight into how melt is extracted from partially molten regions. ``The same basic process of buoyancy-driven melt transport, accompanied by deformation and compaction of the solid matrix'' \citep{shirley1986compaction}, should govern melt transport in partially molten ammonia-rich ice shells. 

Here we develop a one-dimensional model for melt migration in an ammonia-rich ice shell. We model coupled melt migration and thermal evolution during the formation of an ice shell from an initially warm and thin state. Our goal is to understand whether ammonia will freeze into the ice shell as it thickens or whether the ammonia will migrate down into the ocean. We also wish to understand the migration timescale for partial melts and understand the effect of ammonia transport on the thermal evolution of Triton.  

\section{An Ammonia-Water Ice Shell}
The phase diagram of the ammonia-water system is necessary for understanding the initial structure of an ammonia-rich ice shell. 
Figure 1 shows a simplified phase diagram of the ammonia-water system at atmospheric pressure. The liquidus of the system decreases with increasing mass concentration of ammonia from $273.1$ K down to a eutectic of $176.15$ K \citep{kargel1992ammonia,LK2002}. Below the solidus temperature of $176.15$ K (which is independent of ammonia concentration), the stable phases are water ice mixed with ammonia-dihydrate ice. 

In between the solidus and the liquidus temperatures, the stable phases are solid water ice mixed with ammonia-rich fluids, and for a given temperature $T$ and bulk concentration of ammonia $\chi_{NH_3}$, only a fraction of solids will melt. This equilibrium melt volume fraction is calculated as

\begin{equation}
\phi_{eq}=\frac{\chi_{NH_3}}{X_l}
\end{equation}

where $X_l$ is the concentration of ammonia in the liquid. The bulk ammonia concentration is defined as the mass fraction of ammonia in a region (including both the liquids and the solids present). We use the symbol $\chi$ to denote bulk concentration, and $X$ to denote the concentration in the liquid. We calculate $X_l$ as a function of temperature from equations given in \citet{LK2002}. The equilibrium melt fraction is related to the dynamic melt fraction $\phi$ which can change due to melt-solid separation. For reference, all variables are also described in Table 1. 

Figure 2 shows an example of the equilibrium melt fraction plotted as a function of depth in the ice shell for the case of a constant temperature gradient and a bulk concentration of $\chi_{NH_3}= 10\%$ which is constant with depth. Because there is a large temperature range in which partial melt is stable, there may be large partially molten regions in icy bodies whose thickness depends on the thermal gradient and ammonia concentration. Such regions of partial melt may quickly compact as ammonia-rich melt migrates out of the ice shell and into the subsurface ocean, but the rate of compaction depends on many factors, most importantly on the melt fraction. 

Where the melt fraction is high (above some critical melt fraction $\phi>\phi_c$) solids will be suspended in the fluids. Where the melt fraction is lower ($\phi<\phi_c$) solids will be in contact with one another so that they form a porous solid layer (e.g. \citealt{brophy1991composition}). These regions are depicted in figure 2. 

Solid and liquid separation occurs rapidly in the region where crystals are suspended. Water ice crystals with a density $\rho_i=0.92$ g/cm$^3$ will be buoyant with respect to the ammonia-water solution $\rho_l=0.945-1$ g/cm$^3$ \citep{croft1988equation}, and will float upward until they reach the boundary where solids are in contact ($\phi=\phi_c$). The upward velocity of water ice crystals can be described by hindered Stokes settling (e.g. \citealt{suckale2012crystals}). The rate at which solids and liquids separate vertically is defined as the separation flux, $S$, and is a volume flux per area. The separation flux of crystals in this layer is approximated as \citep{solomatov2000fluid}

\begin{equation}
S=0.1\frac{\Delta \rho g d^2}{18 \eta_l} ,
\end{equation}

where $g$ is gravity, $d$ is grain size, $\eta_l$ is the viscosity of the liquid, and $\Delta \rho =80$ kg/m$^3$ is the density contrast between water ice and water.  For $d=1$ mm \citep{BarrMcKinnon2007}, $\eta_l=10^{-3}$ Pa s, and $g=0.78$ m/s$^2$ \citep{burns1986satellites}, parameters appropriate for Triton, the separation flux is $\sim10$ km/year. This shows that the timescale for crystal flotation is much shorter than the timescale for ice shell thickening.

Fluid-crystal separation in the region $\phi<\phi_c$ occurs by porous flow accompanied by compaction of the porous ice shell. The separation flux in this region could be much slower, since the rate of fluid flow can depend on the rate at which the solid matrix can deform and compact. The rate of ice shell compaction depends on the viscosity of the ice shell, which is strongly temperature dependent. To calculate melt migration through a porous ice shell, it is therefor necessary to simultaneously solve for the thermal evolution. 

\section{Methods}
We model the migration of ammonia-rich melt during ice shell formation. We consider an idealized scenario where the ice shell is initially $1$ km thick and cools by conduction. As the ice shell grows, ammonia-rich melt migrates down toward the ocean but the depth of the solidus temperature also migrates downward. If ammonia melt is still present at the depth of the solidus, ammonia will freeze into the ice shell. This problem is similar to the accumulation of the anorthositic crust on the Moon and the potential trapping of interstitial melt in the crust \citep{piskorz2014formation}. It is also very similar (but upside-down) to compacting igneous cumulates at the base of a magma chamber \citep{shirley1986compaction}.  The model described here is used to examine the first $50$ Myr of ice shell evolution. In section 5.1 we use a separate numerical model to examine the long-term thermal evolution of Triton over 4.5 Gyr. 

\subsection{Numerical Model}
We use a one-dimensional finite difference code to simultaneously solve for the thermal evolution and the advection of melt and ammonia. At the top of the ice shell the fluid flux is set to zero and the temperature to $T_s=38$ K, the approximate surface temperature of Triton \citep{burns1986satellites}.  At the base of the ice shell the melt fraction is set to $\phi = \phi_c$ and $T=T_l$, where $T_l$ is the liquidus temperature. We assume a critical melt fraction of $\phi_c=0.5$.

To determine the compaction rate of partially molten regions, we calculate the separation flux $S$ between the liquids and the solids, which can be described by Darcy's law in terms of the compaction viscosity of the solid $\xi$, and the permeability $K$, where
\begin{equation}
S=\frac{\Delta \rho g K}{\eta_l}\bigg{[}(1-\phi)+\frac{1}{\Delta \rho g}\frac{\partial}{\partial z}\bigg{(}\xi\frac{\partial S}{\partial z}\bigg{)}\bigg{]}.
\end{equation}

Here $z$ is vertical distance below the surface, the first term represents flow driven purely by buoyancy and the second term represents flow driven by pressure differences between the solid and the liquid as a result of compaction \citep{sparks1991melt}. 

The rate of change of melt fraction and ammonia concentration in each region are calculated as 

\begin{equation}
\frac{\partial \phi}{\partial t}=-\frac{\partial}{\partial z}\bigg{(}v_l \phi\bigg{)} +\gamma,
\end{equation}

\begin{equation}
\frac{\partial \chi_{NH_3}}{\partial t}=-\frac{\partial}{\partial z}\bigg{(}v_l X_l \phi\bigg{)}.
\end{equation}

where $v_l$ is the velocity of the liquid, and $\gamma$ is the melt production rate. The velocity of the liquid relates to the separation flux as $ v_l=S(1/\phi-1)$. The melt production rate $\gamma$ is calculated in terms of rate of change in temperature and the concentration of ammonia in the liquid (see Supplemental Material).

We assume that heat transport in the ice shell occurs purely by thermal conduction. In our one-dimensional model, fluids and solids are always locally at the same temperature. Solid-state convection of the ice shell could occur if the ice shell is sufficiently thick, however, we focus on a time when the ice shell is thin and cooling solely by conduction. The rate of change of temperature is therefore,

\begin{equation}
\frac{\partial T}{\partial t}=\frac{k}{\rho C_p^*}\frac{\partial^2T}{\partial z^2}
\end{equation}
 
where $k$ is the thermal conductivity of ice, and $C_p^*$ is the effective specific heat of the water ice and liquid mixture. Currently we do not consider a heat source term, such as from tidal heating. The effective specific heat accounts for the heat released due to freezing, where
 
\begin{equation}
C_p^*=C_p\bigg{(}1+\frac{L}{C_p}\frac{\partial \phi}{\partial T}\bigg{)}
\end{equation}

where $L$ is the latent heat of freezing and $\frac{\partial \phi}{\partial T}$ is the change in melt fraction with temperature. Below the solidus temperature the melt fraction abruptly goes to zero as ammonia-dihydrate ice forms. For numerical stability we widen the temperature range over which melt goes to zero to $T_{range} = 176 \pm 2 $ K (see supplemental material). 

The growth rate of the ice shell, $V_H$, is solved for both in terms of the thermal evolution and the compaction rate, such that

\begin{equation}
V_H=V_d - V_c 
\end{equation}

where $V_d$ is the rate of new ice formation at the base and $V_c$ is the rate of ice shell compaction due to the expelling of fluids. The compaction rate is equal to the separation flux at the ice shell-ocean boundary. The rate of new ice formation at the base is calculated such that the latent heat released at the boundary is equal to the heat conducted away from the boundary (see \citealt{turcotte2002geodynamics}). We benchmarked our thermal evolution numerical model against analytical solutions to the Stefan problem.

\subsection{Permeability and Compaction Viscosity}
We assume the permeability of ice is a function of the melt fraction such that

\begin{equation}
K=K_{ref}\phi^m, 
\end{equation}

where $K_{ref}$ is the reference permeability and $m$ is the permeability exponent. Sea ice permeability measurements are well fit by an exponent $m=3$ and the reference permeability in sea ice may range from $K_{ref}=10^{-8}$ to $10^{-10}$ m$^2$ \citep{golden2007thermal}. In our simulations we use $m=3$ and $K_{ref}=10^{-9}$ m$^2$. 

A dramatic decrease in the permeability of sea ice has been observed below melt fractions $\phi<0.05$ due to melt pockets becoming isolated. 
However, in the ammonia-water system, inter-granular melt channels have been observed to form a connected network at temperatures as low as $T=190$ K due to a low dihedral angle $\theta<30^{\circ}$ \citep{goldsby1994structure}. We therefore assume that the ammonia-ice system remains permeable down to the solidus. The possibility remains, however, that the ammonia-ice system could become impermeable at some cut off melt fraction near a few percent. If this were the case, potentially more ammonia-rich melt could become frozen into the ice shell. 

The compaction viscosity, $\xi$, is also strongly dependent on the melt fraction. As the melt fraction becomes smaller, melt channels become increasingly narrow and difficult to squeeze out, so the compaction viscosity is often formulated as \citep{shirley1986compaction}

 \begin{equation}
\xi=\frac{\xi_{ref}}{\phi}.
\end{equation}

The reference compaction viscosity $\xi_{ref}$ is related to the shear viscosity of the solid $\eta_s$. The shear viscosity of ice is strongly temperature dependent, however, partial melt can also affect the shear viscosity. For partially molten ammonia-rich water ice, the shear viscosity decreases about an order of magnitude for each increase in melt fraction of $0.1$ and the viscosity can be approximated as $\eta_s \sim \exp(-B\phi)$, where $B\approx22$ is an empirical constant \citep{arakawa1994effective}. 
 
We adopt a compaction viscosity that is a function of both temperature and melt fraction, 

\begin{equation}
\xi=\frac{\xi_{ref}}{\phi}\exp\bigg{(}\frac{Q}{R}(\frac{1}{T}-\frac{1}{273.1})\bigg{)}\exp{(-22\phi)},
\end{equation}

where $R$ is the gas constant and $Q$ is the activation energy for viscous creep. For pure water ice, the activation energy ranges from $49 - 60$ kJ/mol, depending on the deformation mechanism \citep{GoldsbyKohlstedt2001}. However, \citet{arakawa1994effective} measured the viscosity of partially molten ammonia-rich ice and found an activation energy of $33$ kJ/mol. For our simulations we select $Q=33$ kJ/mol, and as is demonstrated later, the temperature dependence of the compaction viscosity strongly influences melt migration patterns.

The density of the liquid changes with ammonia concentration \citep{croft1988equation}. If the composition of the liquid stays in  equilibrium with the surrounding solid ice, the density of the fluid can be expressed as a function of temperature. We fit a quadratic function to density measurements by \citet{croft1988equation},

\begin{equation}
\rho_l = 1.0 + a(T-273.1)+b(T-273.1)^2 \frac{g}{cm^3}
\end{equation}

where $a=1.7\times10^{-3}$ and $b=1.3\times10^{-5}$. The reduced density contrast at low temperatures significantly reduces the rate of melt migration. 

The viscosity of the fluid also changes drastically as a function of ammonia-concentration and temperature.
 The viscosity of ammonia-dihydrate liquid at the eutectic temperature is approximately $\eta_l=5$ Pa s \citep{kargel1991rheological}, nearly four orders of magnitude more viscous than pure water at $T=273.1$ K. We fit an exponential law to the viscosity data by \citet{kargel1991rheological}, to account for how the fluid viscosity changes as function of temperature, where

\begin{equation}
\eta_l= \eta_{l,ref}\exp{\bigg{(} p_1 (T^2-273.1^2) + p_2 (T-273.1)\bigg{)}}
\end{equation}

where $\eta_{l,ref}=10^{-3}$ Pa s is the reference fluid viscosity, and $p_1=5.85\times10^{-4}$, $p_2=-0.3445$ are empirical constants and temperature is in Kelvin.
 
As the temperature approaches the solidus, the combined effects of higher compaction viscosity, higher fluid viscosity and reduced density contrast make segregation of ammonia-rich melts difficult.

\subsection{Non-Dimensionalization}
We compute our solutions in non-dimensional form. Non-dimensional solutions are useful for extrapolating our results to bodies with different physical parameters, such as gravity or reference compaction viscosity. We normalize length by a reference compaction length

\begin{equation}
L_{ref}=\sqrt{\frac{K_{ref}\xi_{ref}}{\eta_{l,ref}}}.
\end{equation}

We use a reference compaction viscosity of $\xi_{ref}=10^{14}$ Pa s, appropriate for ice near the melting temperature with a grain size of $\sim1$ mm \citep{GoldsbyKohlstedt2001}. Thus the reference compaction length scale for the ice shell $L_{ref}\approx10$ km. 

Velocity is non-dimensionalized by the reference separation flux,

\begin{equation}
S_{ref}=\Delta\rho_{ref} g K_{ref}/\eta_{l,ref}, 
\end{equation}

where $\Delta\rho_{ref}$ is the density contrast between pure water and solid ice at $T=273.1$ K. 
For the examples shown below, we use $S_{ref}=6.24\times10^{-5}$ m/s. Time $t$ is normalized by $L_{ref}/S_{ref}$. The non-dimensional ice shell thickening rate is referred to as $\bar{V}$. 

\subsection{Benchmarking}
Our numerical model is similar to \citet{shirley1986compaction}, who investigated the solidification and compaction of terrestrial magma chambers. His model examined a thickening layer of partially molten igneous cumulates being deposited at the base of a cooling magma chamber. For melt to be extracted from this layer, the solid-matrix must deform and compact. His model identified two distinct modes of compaction. If the deposition rate of new crystals at the top of the layer is rapid, the majority of the layer stays near the value $\phi_c$. If the deposition rate is slow, melt is buoyantly expelled over most of the layer except in a boundary layer near the solidification front where melt migration is limited by the compaction rate. 

A benchmark to \citet{shirley1986compaction}, showing these two compaction modes, is shown in figure 3 for two different constant deposition rates. In this case the solidification front propagates downward, to match the geometry of ice shell thickening. When the non-dimensional thickening rate is high, the ice shell growth rate is fast compared to the compaction rate, and melt cannot be expelled from the thickening shell. This example is relevant to when the ice shell is thin and rapidly cooling. To demonstrate our treatment of melt migration in an ice shell, we also benchmarked our model to \citet{kalousova2014ice}, who modeled the propagation of melt through Europa's ice shell (see supplemental material).

\section{Results}
\subsection{Melt Migration in Constant Thickness Ice Shell}
Before modeling melt migration in a growing ice shell, we first consider a 10 km thick ice shell of constant thickness, in which a constant temperature gradient is fixed with time. The ice shell is composed of $10$ wt.$\%$ ammonia and the initial melt fraction is set to the equilibrium melt fraction (the predicted volume fraction of melt for a given bulk ammonia concentration and temperature). Partial melt is initially present below depths of $6.5$ km, where the temperatures are above the solidus. Figure 4 shows how melt evolves over time in this region. The melt separation flux is much higher at warmer temperatures, where the compaction viscosity and fluid viscosity is lower. Melt in the warmer portions of the ice shell drains out first, leaving behind ammonia-rich melt in colder portions of the ice shell, which takes much longer to drain out. Where melt has drained out, the ice shell has become depleted in ammonia and is nearly pure water ice. However, just below the solidus boundary, a melt fraction of approximately $10\%$ remains in the ice shell even after $300$ kyr of evolution. 

The temperature dependence of the fluid viscosity and compaction viscosity drastically affects melt migration in the ice shell. Figure 5 shows how the average melt fraction in the ice shell evolves over time, with and without temperature-dependent viscosity. Similar to the example shown in figure 4, we consider a 10 km thick ice shell of constant thickness and composed of $10\%$ ammonia.  Including a temperature-dependent compaction viscosity, versus a constant compaction viscosity of $\xi=10^{14}$ Pa s, increases the timescale for melt migration by two orders of magnitude. For instance, with no temperature dependence (blue line in Figure 5) the average melt fraction drops to $\phi=0.05$ in $\sim100$ yr, but with a temperature dependent viscosity it takes $10^4$ yr for the average melt fraction to reach this value (red line Figure 5). Including a temperature-dependent fluid viscosity further increases the melt migration timescale (black line in Figure 5). It is thus essential to consider temperature-dependent viscosities when modeling the migration of low temperature eutectic melts. 

\subsection{Melt Migration in a Growing Ice Shell}
We also model melt migration in a thickening ammonia-rich ice shell. Our ice shell is initially $1$ km thick, has an ammonia concentration of $\chi_{NH_3} = 10\%$ and an initially linear temperature profile. The initial melt fraction is calculated assuming thermal equilibrium. 

Figure 6 shows how the temperature, melt fraction and ammonia concentration in the ice shell evolve over time. As the ice shell thickens the solidus front propagates downward, causing ammonia-rich melt to freeze into the upper $\sim8$ km of the ice shell in concentrations of $\chi_{NH_3}>0.01$. The latent heat released from freezing slows the propagation of the solidus front but not enough to prevent ammonia from freezing into the ice shell. A compaction boundary layer develops at the base of the ice shell that is approximately $100$ m thick. Above this compaction layer, a region of relatively constant melt fraction develops, similar to the melt structures calculated by \citet{shirley1986compaction}.

The compaction viscosity at the solidus temperature is approximately $\xi \approx 10^{20}$ Pa s, making it difficult for melt to be extracted if melt is present at this temperature. Additionally, higher fluid viscosity and reduced melt density (which reduces the liquid to solid density contrast) act to slow the rate of compaction at low temperatures. In warmer regions of the ice shell, melt is extracted more efficiently, and ammonia begins to be removed from the ice shell. The concentration of ammonia, which has frozen into the ice shell, decreases with depth (Figure 6). In this example, the ice shell above $8$ km depth contains $\chi_{NH_3}>0.01$ ammonia, and the ice shell below is depleted in ammonia and is composed of relatively pure water ice below this depth. Figure 7 shows the melt fraction and ammonia concentration in the ice shell for the same simulation, with time on the horizontal axis. When the ice shell is cooling rapidly enough, the freezing of ammonia-rich melts at the solidus front causes a significant amount of ammonia to remain frozen in the ice shell. 

\section{Discussion}
During ice shell formation, we find that eutectic melts of ammonia can become frozen in the ice shell when the cooling rate is faster than the compaction rate. At low temperatures, the high compaction viscosity, high fluid viscosity and reduced density contrast between the melt and solid, significantly reduce the compaction rate and increase the propensity for ammonia-rich melts to freeze. These temperature-dependent effects are particularly important for ammonia-rich melts, where the solidus is nearly $\sim100$ K below the liquidus. 

In our example calculation, ammonia becomes frozen into the upper $\sim8$ km of the ice shell, and the ice shell below this depth is relatively pure water ice. However, the results shown in figures $4$ to $7$ should only be taken as example calculations with nominal values for compaction viscosity and permeability. If the reference compaction viscosity is higher, the thickness of the region where ammonia becomes trapped in the ice shell will be increased. For example, if the reference compaction viscosity were two orders of magnitude larger at $\xi_{ref} = 10^{16}$ Pa s, ammonia could potentially become trapped in the upper $\sim30$ km of the ice shell (see supplemental material S4). We also find that the amount of ammonia that freezes into the ice shell is insensitive to the chosen value of the reference permeability, suggesting that melt migration is rate-limited by the compaction rate of the porous ice shell. 

We only consider a simplified scenario where the ice shell cools from a very thin initial state. This approach allowed us to examine the coupled dynamics of melt migration and thermal evolution in the ice shell at a time when the cooling timescale is comparable to the compaction timescale. However, the formation of ice shells on icy bodies is likely considerably more complex, and ice shell formation may occur concurrently with many other processes such as accretion, impact cratering, differentiation and tidal heating. However our general result can still be applied to bodies with more complex geological and thermal histories, which is that low-temperature eutectic melts can become frozen into the ice shell if the cooling rate is faster than the compaction rate, and that the high compaction viscosity of ice at $T=176$ K increases the propensity for ammonia to become frozen into the ice shell. 

The freezing of volatiles into the near-surface is significant. If the volatile-rich near-surface of an icy satellite is reheated, such as during a period of tidal heating or by the onset of solid-state convection in the ice shell, eutectic melts could be generated. It is thought that tidal heating or convective upwellings may generate eutectic melting beneath chaos terrain on Europa \citep{sotin2002europa}, and perhaps beneath double ridges on Europa \citep{nimmo2002strike}. \citet{kalousova2014ice} modeled melt migration through Europa's ice shell and found that near-surface melts would migrate down to the subsurface ocean in approximately $10^4$ years, if the ice shell were temperate, and if the melts were pure water. If eutectic melts of ammonia-water are generated, the high compaction viscosity of ice and high fluid viscosity at low temperatures will reduce the rate of melt migration. The reduced melt migration rate of low-temperature eutectic melts could allow them to remain near the surface for much longer than pure water melts. Ammonia may not be present in Europa's ice shell, but other contaminants with low eutectic temperatures could be present, such as hydrated-chloride salts or sulfuric acid \citep{kargel2000europa}. 

Further studies are needed to understand the formation and evolution of near-surface eutectic melts on tidally heated icy satellites. Future models could examine more complex chemical systems, melt transport through fractures and Rayleigh-Taylor instabilities, as well as the influence of topographically induced pressure gradients (c.f., \citealt{showman2004resurfacing, neveu2015prerequisites, kalousova2016water, quick2016heat}). Incorporating eutectic melting into advanced two and three dimensional models could significantly advance our understanding of melt migration on icy satellites. 

Although our work shows that volatile trapping can occur near the surface, we also find the majority of ammonia-rich melt is transported into the subsurface ocean. This suggests that subsurface oceans will become enriched in volatiles over time as the ice shell continues to thicken and exclude ammonia. We explore the significance of this result below with respect to the geological evolution of Triton.

\subsection{Thermal Evolution of Triton}
Triton is the only large satellite of Neptune. Much of the surface is covered in ``cantaloupe terrain'', which some authors have suggested has a cryovolcanic origin \citep{croft1995geology}. The dearth of impact craters on Triton suggests a relatively young surface \citep{stern2000triton}. Triton has a retrograde orbit around Neptune, implying it was likely gravitationally captured \citep{agnor2006neptune}. Just after orbital capture, Triton would have had a very high eccentricity, causing it to experience intense tidal heating \citep{ross1990coupled}. The magnitude of tidal heating was likely large enough to cause differentiation and melt a majority of the ice shell \citep{ross1990coupled}. Significant tidal dissipation causes orbital eccentricity to dampen, and Triton's current orbital eccentricity is near zero \citep{burns1986satellites}. The timescale for the circularization of Triton's orbit is estimated to be between $10^5$ to $10^8$ yr, \citep{ross1990coupled,cuk2005constraints}. After Triton's orbit circularized, tidal heating would become insignificant compared to the radiogenic heat flux from the core. We note that long-term tidal heating could still be significant if orbital circularization proceeded more slowly \citep{gaeman2012sustainability}. It has also been suggested that obliquity tides could cause significant energy dissipation in the Triton's subsurface ocean \citep{chen2014tidal}, however dissipation in icy satellite oceans is still poorly understood and ocean tidal dissipation is not included in our model.

We model the thermal evolution of Triton after its orbit has circularized and demonstrate the effect of increasing ammonia concentration in the ocean as the ice shell thickens. We use a 1-dimensional thermal conduction model similar to that used by \citet{hammond2016recent}. The H$_2$O layer thickness is $330$ km and the silicate core has a radius of $R_c=1023$ km. Both the core and the ice shell transmit heat purely by conduction and the ocean is assumed to be vigorously convecting such that the temperature in the ocean is constant with depth. Solid-state convection in Triton's ice shell may be possible after the ice shell becomes sufficiently thick, which would cause Triton's subsurface ocean to freeze more rapidly. Here we only explore cases where the ice shell is not convecting, and demonstrate the effect of gradually increasing ammonia in Triton's subsurface ocean.

Radioactive isotopes in the silicate core produce heating and we assume the core has the same concentration of radioactive isotopes as CI chondrites \citep{lodders2003solar}. The abundance of radioactive isotopes can be found in Table 3 of \citet{robuchon2011thermal}. We use only 400 ppb of potassium-40 in the silicate core due to the susceptibility of potassium to volatile depletion during accretion \citep{humayun1995potassium}. We begin the simulations $30$ Myr after CAI formation so short lived radioisotopes do not have a significant effect. The thermal conductivity in the core is $k=3$ W/mK and the thermal conductivity in the ice shell $k_i=0.48+488/T$ W/mK 
\citep{petrenko1999physics}. We assume a simple initial condition, using a uniform internal temperature with an ice shell $30$ km thick and an initial ammonia mass concentration in the ocean $\chi_0=5\%$.

To account for the process of ammonia segregating from the ice shell into the ocean as the ice shell forms, we increase the ammonia concentration in the ocean as

\begin{equation}
\chi_{NH_3}=\frac{\chi_0}{1-V_{Shell}/(V_{H_20})}
\end{equation}

where $V_{Shell}$ is the volume of the ice shell and $V_{H_20}$ is the total volume of the $H_20$ layer (ocean and ice shell). The equations for the melting temperature as a function of pressure and ammonia concentration are taken from \citet{LK2002}.

Figure 8 shows how the temperature evolves in Triton's ice shell with and without accounting for the increasing concentration of ammonia. After $4.5$ Gyr of cooling the ice shell is $270$ km thick and there is a $60$ km thick ocean. The ammonia concentration in the ocean has increased to $\chi_{NH_3}=25\%$. Without accounting for the increase in concentration, the ocean would only be $10$ km thick. Thus accounting for the segregation of ammonia-rich melts makes a significant difference in Triton's thermal and geological evolution. The thermal evolution of subsurface ocean on other icy bodies, such as Pluto, may also be strongly affected by gradually increasing ammonia concentration.

As Triton's subsurface ocean becomes more enriched in ammonia its density decreases. This raises the possibility that the ammonia-rich ocean could become buoyant and lead to cryovolcanic resurfacing. After the ice shell thickens and the ocean reaches a critical ammonia fraction and density, it could become unstable and lead to an overturn. Although the density of the ocean is still higher than the ice shell, the reduced ocean density, combined with ocean pressurization due to ice shell thickening \citep{manga2007pressurized}, could potentially drive upward migration of melt. Perhaps geologically recently, the buoyant ocean could have risen towards the surface and lead to catastrophic resurfacing. This would be an interesting way to explain why Triton's surface appears to be extremely young and possibly cryovolcanically resurfaced \citep{croft1995geology,stern2000triton}.

\section{Conclusion}
To understand the possible distribution of ammonia in icy satellite ice shells and subsurface oceans, we modeled ammonia-melt migration during the solidification of an ice shell from an initially warm and thin state. We find that ammonia can be frozen into the upper few kilometers of the ice shell, owing to the high compaction viscosity of ice at the solidus temperature. The freezing of volatiles in the near-surface suggests that near-surface eutectic melts could be generated upon reheating of the ice shell by tidal dissipation or convection. 

As the ice shell thickening rate slows, compaction becomes efficient and nearly all the ammonia-rich melt migrates downward into the ocean. The progressive enrichment of ammonia in the subsurface ocean decreases the density and freezing temperature. This makes a significant difference in the thermal evolution of Triton and can help prevent Triton's ocean from freezing, increasing the final ocean thickness by up to 50 km. An initial ammonia concentration of $5\%$ in the ocean can increase to $25\%$ after the ice shell thickens and the reduced density of the ammonia-rich ocean increases the likelihood of cryovolcanic resurfacing.

\section{Acknowledgements}
Author Hammond acknowledges support by NASA NESSF grant NNX13AN99H. Author Barr acknowledges support from NASA OPR NNX09AP30G.The authors also wish to thank Charles-Edouard Boukar\'{e} and Stephen Parman for helpful feedback on this work. All data necessary to reproduce this work can be found at https://github.com/nphammon/Ammonia-water-ice-shell-.


\clearpage
\renewcommand{\baselinestretch}{1}
\small
\begin{center}
\begin{longtable}{| c | c | c | c |}
\caption{Symbols and quantities used in this paper.} \label{symboltable} \\ 
\hline
\textbf{Symbol} & \textbf{Variable Name} & \textbf{Value/Units} & \textbf{Reference} \\
\hline
\endfirsthead
\multicolumn{4}{c}%
{\tablename\ \thetable\ -- \textit{Continued from previous page}} \\
\hline
\textbf{Symbol} & \textbf{Variable Name} & \textbf{Value/Units} & \textbf{Reference} \\
\hline
\endhead
\hline \multicolumn{4}{r}{\textit{Continued on next page}} \\
\endfoot
\hline
\endlastfoot
- & \underline{\textit{Melt Migration}} & & \\
$d$ & Grain size & 1 mm & \citet{BarrMcKinnon2007}\\
$\gamma$ & Melt production rate & /s & \\
$g$ & Triton surface gravity & 0.78 m/s$^2$ & \citet{burns1986satellites}\\
$K_{ref}$ & Reference permeability & $10^{-9}$ m$^2$ & \citet{golden2007thermal}\\
$K$ & Permeability & m$^2$ &  \\
$m$ & Melt permeability exponent & 3 & \citet{golden2007thermal} \\
$\phi_{eq}$ & Equilibrium melt fraction & $0-1$ &  \\
$\phi$ & Dynamic Melt fraction & $0-1$ &  \\
$\phi_c$ & Critical melt fraction  & $0.5$ &  \\

$\eta_{l,ref}$ & Viscosity of fluid at $T=273.1$, & $10^{-3}$ Pa s & \citet{kargel1991rheological}\\

$\xi$ & Compaction viscosity & $10^{13}$ - $10^{20}$ Pa s & \\
$\xi_{ref}$ & Reference compaction viscosity &$10^{14}$ Pa s & \\
$X_l$ & Concentration of ammonia in the fluid & 0 - 0.36 &  \\
$\chi_{NH_3}$ & Mass fraction of ammonia & $0-1$ &  \\
 $v_l$ & Velocity of the fluid& m/s & \\
 $S$ & Separation flux & m/s &  \\ 
  & & & \\
 & \underline{\textit{Thermal Evolution}} & & \\
$C_p$ & Specific heat of ice & 2.05 kJ/kgK & \citet{petrenko1999physics}\\
$C_p^*$ & Effective specific heat &  &   \\
$k$ & Thermal conductivity of ice & 0.48+488/T W/mK & \citet{petrenko1999physics}\\
$L$ & Latent heat of fusion for ice &330 kJ/kg & \citet{petrenko1999physics}\\
$\rho_i$ & Density of ice & 920 kg/m$^3$ & \\
$\rho_l$ & Density of fluid & 946-1000 kg/m$^3$ & \citet{croft1988equation}\\
$\Delta\rho$ & Density contrast between fluid and ice & 26-80 kg/m$^3$ & \\ 

$Q$ & Activation energy for creep & 33 kJ/mol & \citet{arakawa1994effective}\\
$R$ & Gas constant & 8.31 J/(mol K) & \\
$T$ & Temperature & K & \\
$T_s$ & Surface temperature & 38 K & \citet{burns1986satellites} \\
$T_l$ & Liquidus temperature & 176-273 K & \citet{LK2002}\\
$\chi_0$ & Initial ammonia concentration  & 0.05 & \\
 & & & \\
 - & \underline{\textit{Dimensions}} & & \\
 $H$ & Ice shell thickness with time & m &  \\
 $L_{ref}$ & Reference length & 10$^4$ m & \\
 $S_{ref}$ & Reference velocity & $6.24\times10^{-5}$ m/s & \\
 $t$ & Time & s & \\
 $t_{ref}$ & Reference time & $1.6\times10^8$ s & \\
$V_H$ & Ice shell growth rate & m/s & \\
$V_c$ & Ice shell compaction rate& m/s & \\
$V_d$ & Rate of new ice formation & m/s & \\
 $z$ & Depth below surface & m & \\
\end{longtable}
\end{center}

\clearpage
\begin{figure}
\centerline{\includegraphics[width=26pc]{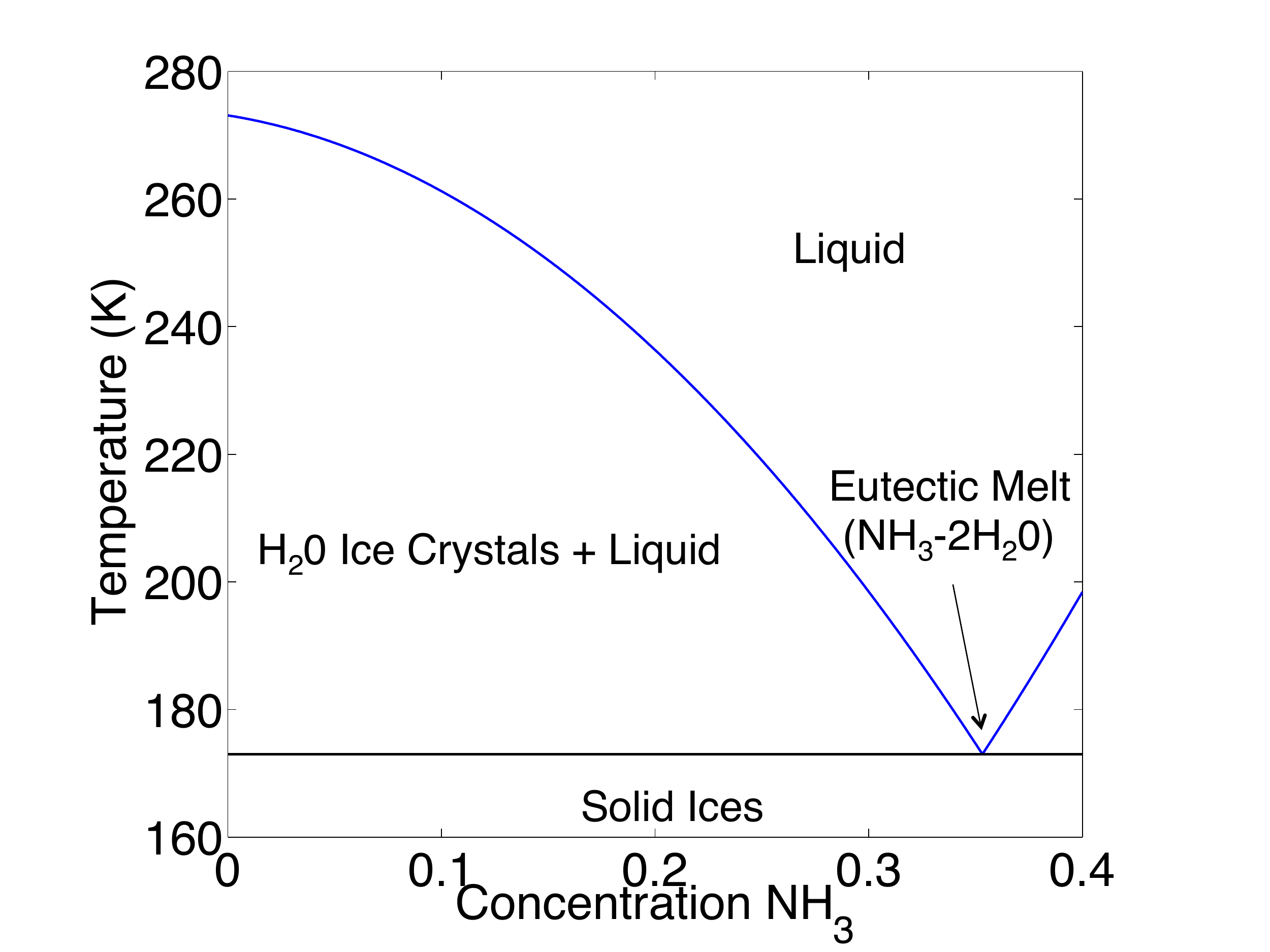}}
\caption[Ammonia-water phase diagram and melt fraction]{The ammonia-water phase diagram at atmospheric pressure, from experiments by \citep{LK2002}. The blue line shows the liquidus as a function of the mass concentration of ammonia. The black solid line shows the solidus.}
\end{figure}

\clearpage
 \begin{figure}
\centerline{\includegraphics[width=26pc]{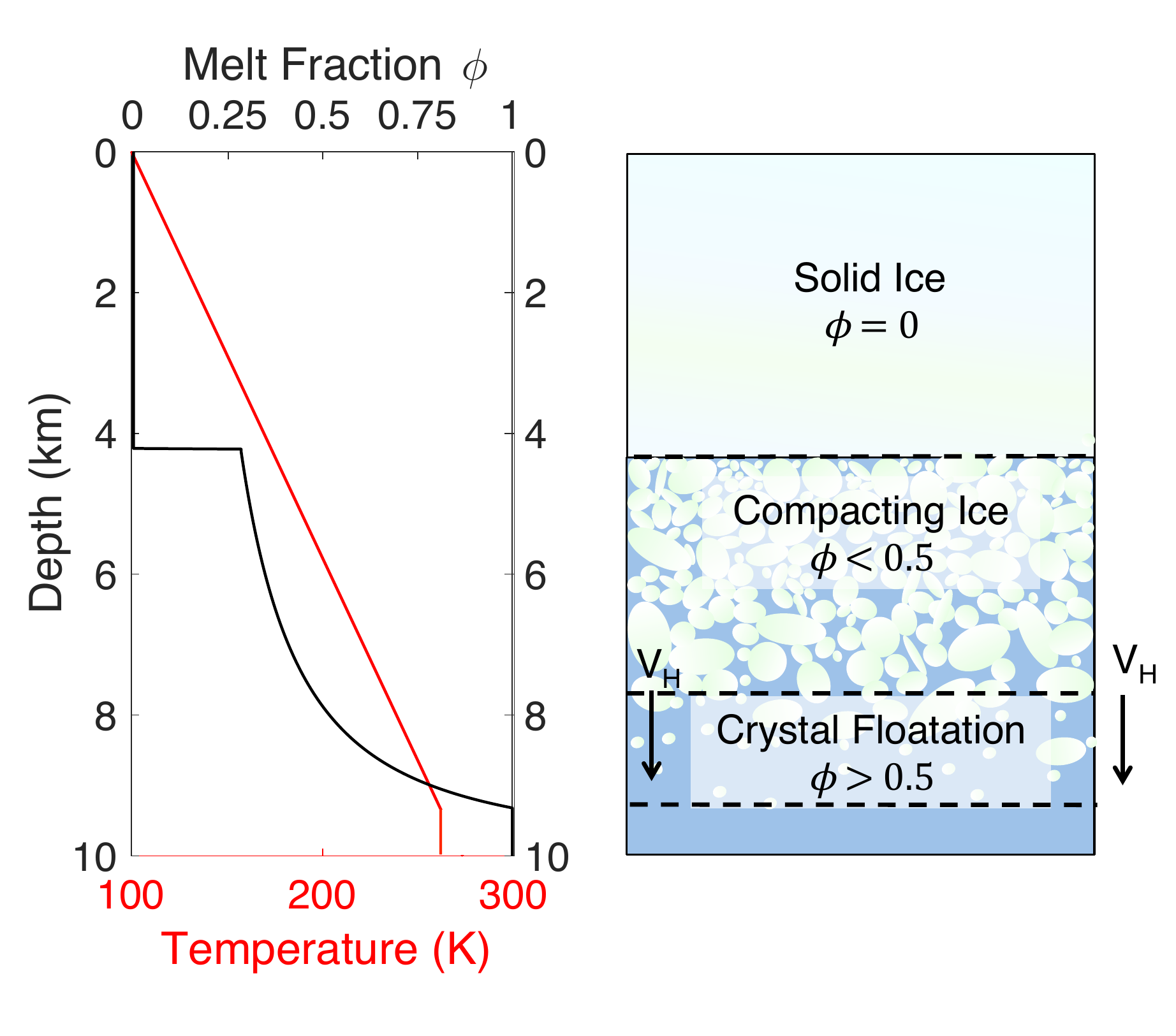}}
 \caption[Melt Fraction Example]{Example of the structure of an ice shell $10$ km thick with a bulk ammonia concentration of $\chi_0=0.1$. The left panel shows the temperature (red line) and equilibrium melt fraction $\phi$ (black line) with depth. The right panel illustrates the regions of the ice shell that are completely solid, compacting and where crystals float toward the base of the ice shell. The rate at which the ice shell thickens $V_H$ is also illustrated.}
 \end{figure}

\clearpage
\begin{figure}
\centerline{\includegraphics[width=26pc]{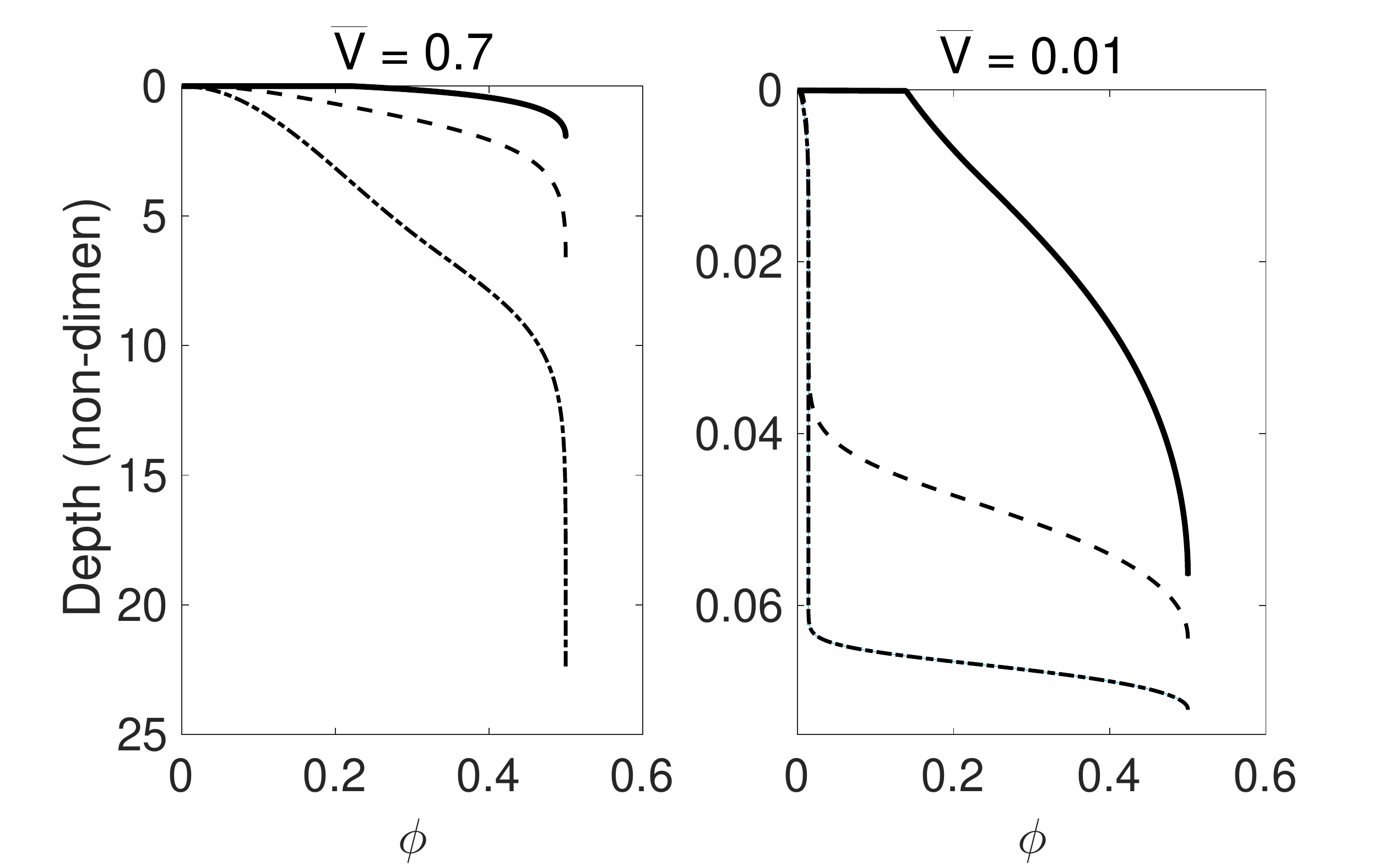}}
\caption[MeltDepthShirley]{A benchmark to \citet{shirley1986compaction} showing two end member cases of compaction in a thickening shell. The melt fraction for three different times are shown in each panel, at non-dimensional times $\bar{t}=2$, $\bar{t}=10$ and $\bar{t}=30$. On the left the shell growth rate is fast compared to the compaction rate, with a constant non-dimensional ice shell growth rate $\bar{V}=0.7$. This causes the average melt fraction in the growing shell to be high. On the right, the compaction rate is fast compared to the ice shell growth rate, with a constant $\bar{V}=0.01$, causing most of the melt to be expelled from the thickening shell. Thermal effects and ammonia concentration are not included in this model. 
} 
\end{figure}

\clearpage
\begin{figure}
\centerline{\includegraphics[width=22pc]{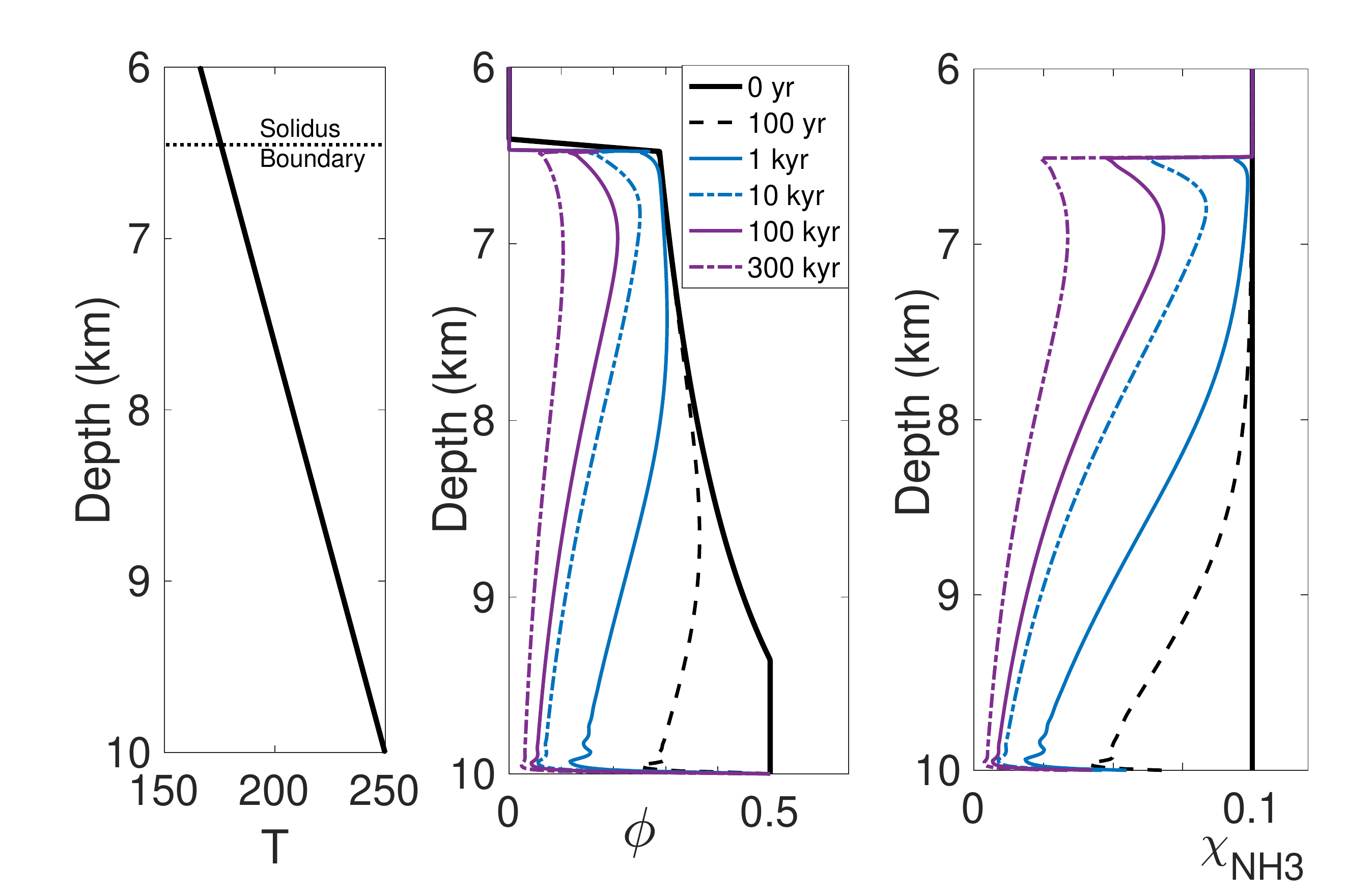}}
\caption[PhiCevolution]{Melt fraction evolution in a constant 10 km thick ice shell with a bulk ammonia concentration of $10\%$. The plot is focused on the region where partial melt is initially present, at depths below $6.5$ km. (left) Temperature with depth in the ice shell. There is a constant temperature gradient throughout the ice shell, with a surface temperature of $38$ K and temperature at the base of the shell of $250$ K. (center) Various solid and dashed lines show the melt profile with depth at different times. (right) Mass fraction of ammonia in the ice shell at different times. 
} 
\end{figure} 

\clearpage
\begin{figure}
\centerline{\includegraphics[width=22pc]{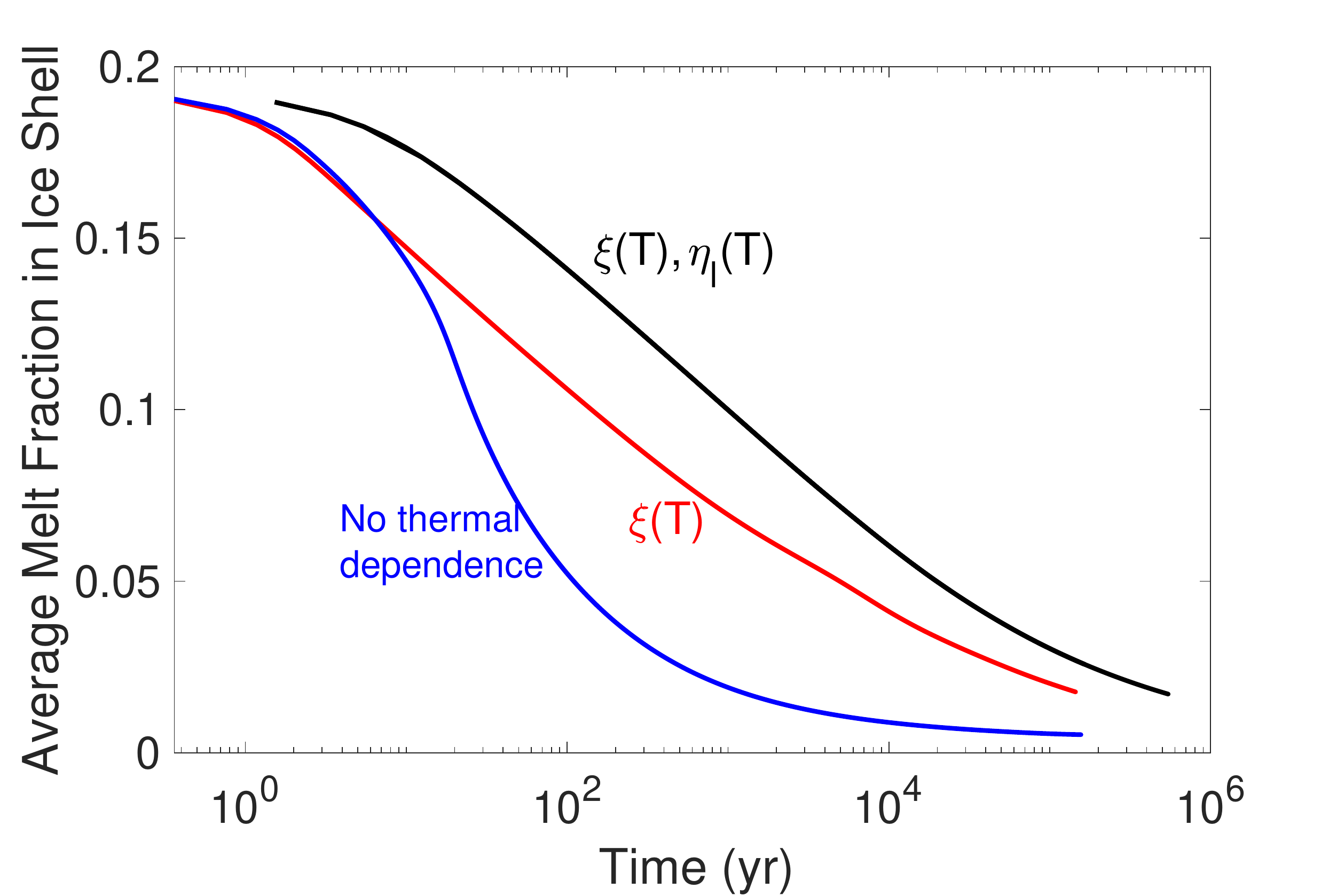}}
\caption[MeltCompare]{The average melt fraction in a 10 km thick ice shell over time. Ice shell thickness and temperature are held constant, as in figure 4. The blue line shows the melt fraction for the case of a constant compaction viscosity of $\xi=10^{14}$ Pa s, and a constant fluid viscosity of $\eta_l=10^{-3}$ Pa s. The red line shows the case of a temperature-dependent compaction viscosity and a constant fluid viscosity, and the black line shows the case where both the compaction viscosity and fluid viscosity are temperature dependent. 
}
\end{figure}

\clearpage
\begin{figure}
\centerline{\includegraphics[width=22pc]{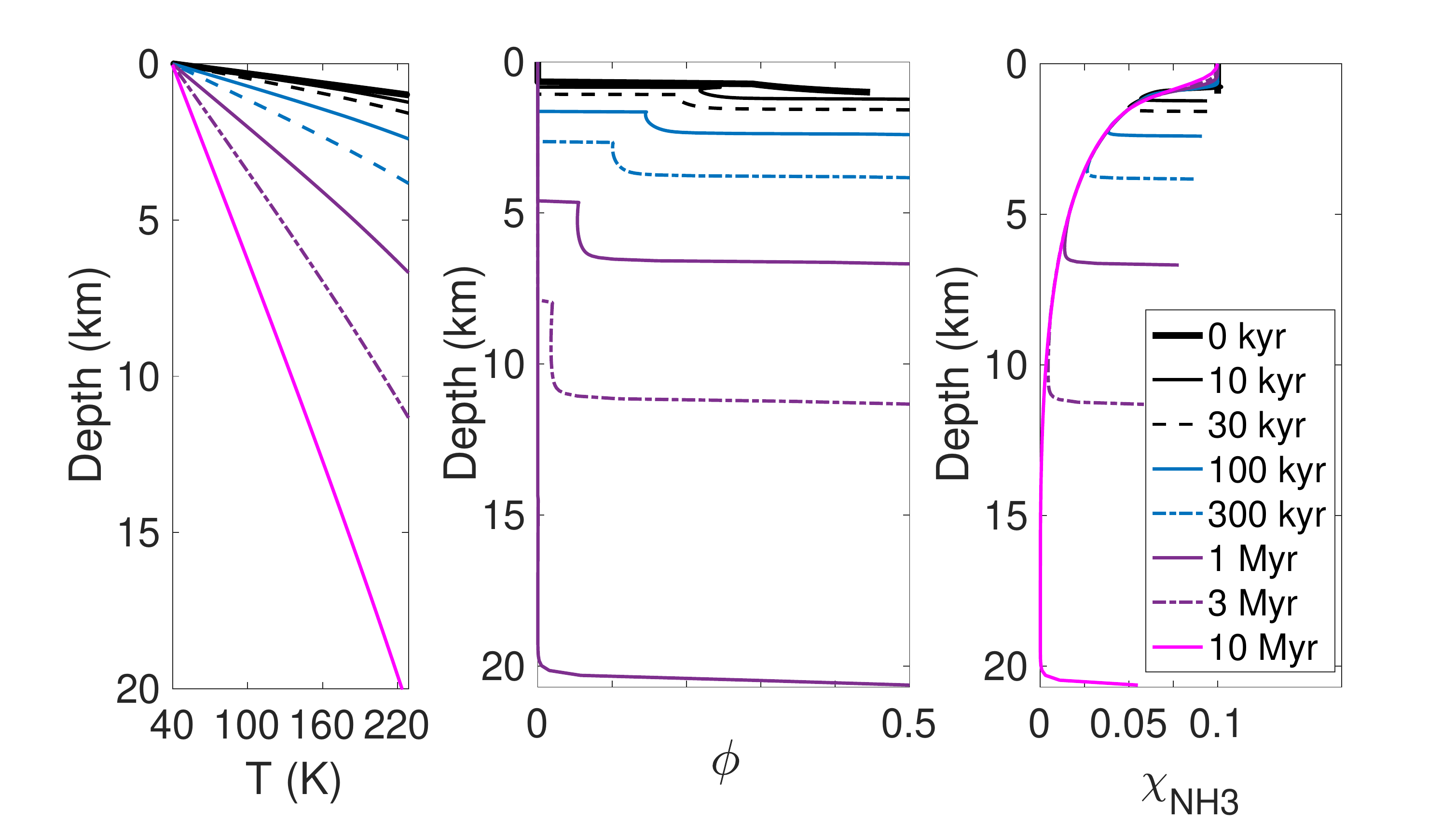}}
\caption[PhiCevolution]{(left) Temperature evolution in a thickening ice shell above an ammonia-rich ocean, with an initial ammonia concentration of 0.1. Various solid and dashed lines show the temperature with depth at different times. (center) Melt fraction in a thickening ice shell. The melt fraction grows sharply to 0.5 near the ice shell-ocean boundary, which propagates downward overtime. (right) Concentration of ammonia in a thickening ice shell. As the growth rate of the ice shell slows, melt is more efficiently extracted and the ice shell becomes depleted in ammonia. 
} 
\end{figure} 

\clearpage
\begin{figure}
\centerline{\includegraphics[width=26pc]{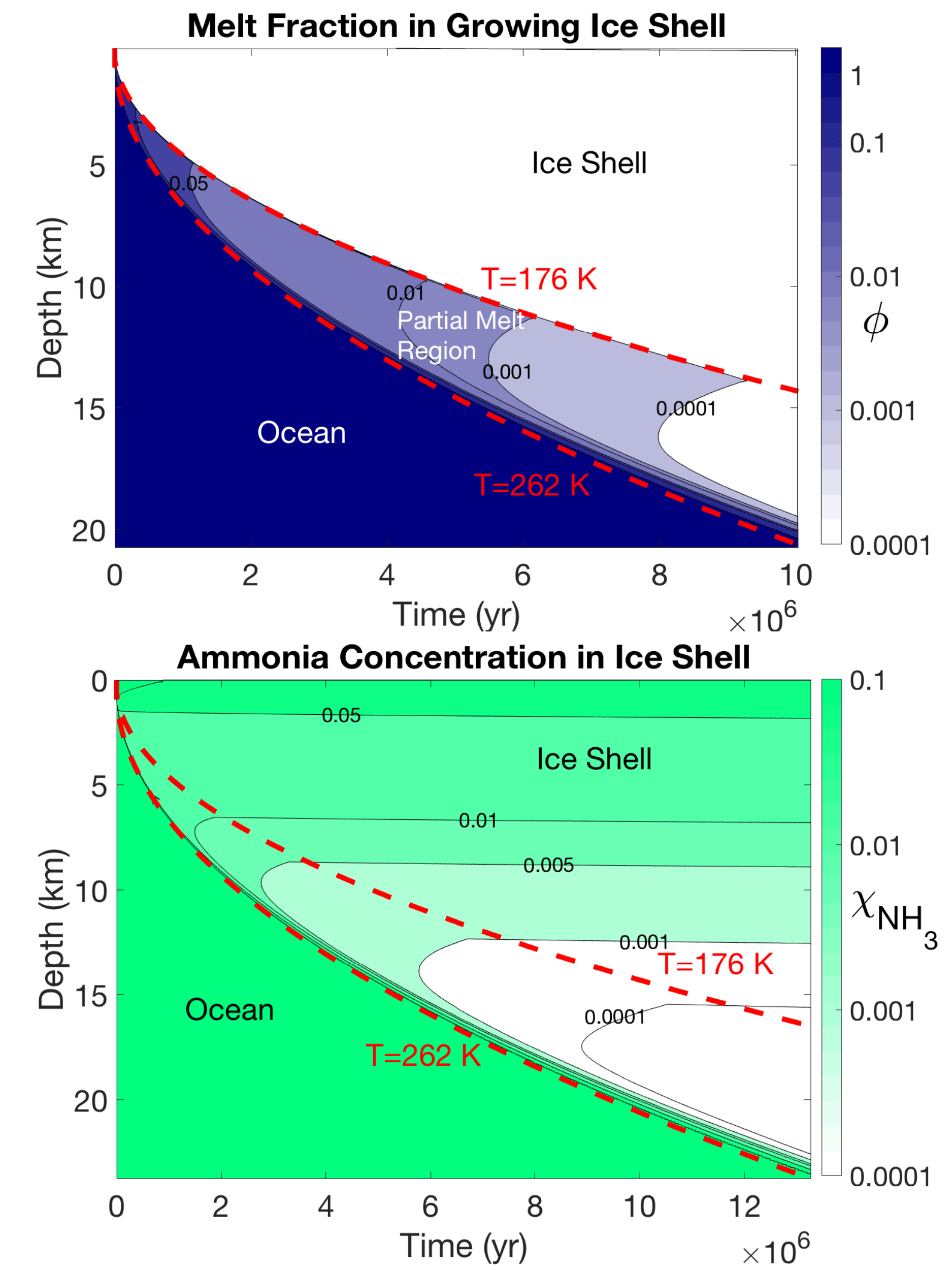}}
\caption[IceShellAmmonia]{The evolution of melt (top plot) and ammonia concentration (bottom plot) in the thickening ice shell. The x-axis is time. The ice shell grows from 1 km to $\sim20$ km thick. Red dashed lines show temperature contours at $T=176$ K and $T=262$ K. (top) Melt fraction in the ice shell over time. Blue filled in contours show melt fraction in the ice shell. (bottom) Ammonia concentration in ice shell and ocean over time. Green filled contours show ammonia concentration with depth and time.} 
\end{figure}

\clearpage
 \begin{figure}
\centerline{\includegraphics[width=26pc]{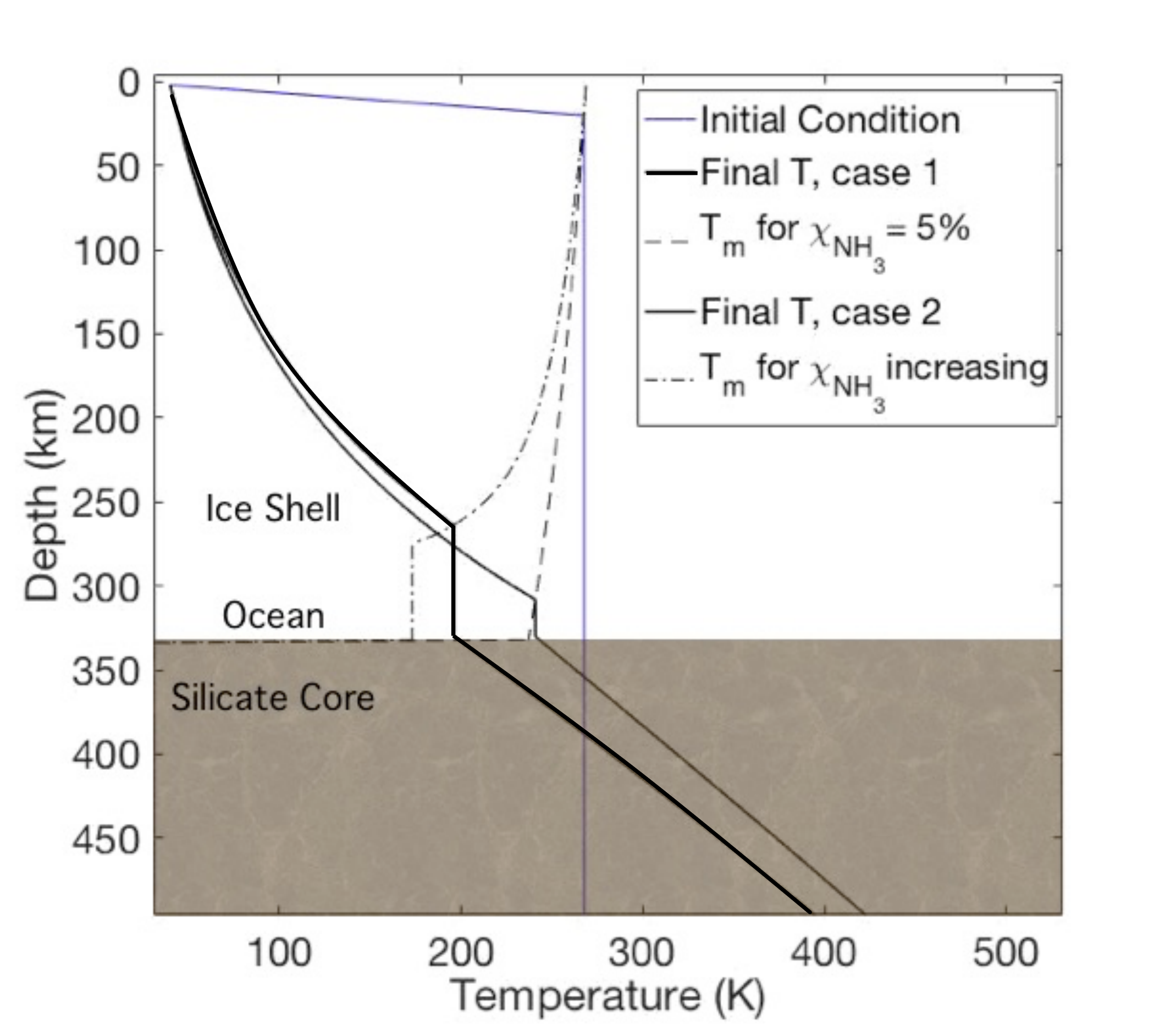}}
\caption[An example of the thermal evolution of Triton with increasing concentration of ammonia in the ocean]{Examples of the thermal evolution of Triton showing the temperature in the ice shell and the upper portion of the silicate core.The blue line shows the initial temperature profile. The solid black lines show the temperature after $4.5$ Gyr of cooling.  The bold solid black line shows case 1, where ammonia concentration increases as ocean volume decreases. The thin solid black line shows the final temperature for case 2, where ammonia concentration in the ocean is held constant. The dashed line shows the melting temperature assuming $5\%$ ammonia in the ice shell and ocean. The dashed-dotted line shows the melting temperature assuming that ammonia becomes concentrated in the ocean over time.}
 \end{figure}

\end{document}